\documentclass{article}
\usepackage{spconf,amsmath,graphicx}
\usepackage{amsmath,graphicx,multirow,array,tikz,subfig,bm,amsfonts,setspace}
\newcommand{\PreserveBackslash}[1]{\let\temp=\\#1\let\\=\temp}
\newcolumntype{C}[1]{>{\PreserveBackslash\centering}p{#1}}
\newcolumntype{R}[1]{>{\PreserveBackslash\raggedleft}p{#1}}
\newcolumntype{L}[1]{>{\PreserveBackslash\raggedright}p{#1}}

\title{MULTISTREAM CNN FOR ROBUST ACOUSTIC MODELING}
\name{Kyu J. Han$^1$, Jing Pan$^1$, Venkata Krishna Naveen Tadala$^2$, Tao Ma$^1$ and Dan Povey$^3$\thanks{Codes are available at https://github.com/asappresearch/multistream-cnn.}}
\address{
  $^1$ASAPP, Mountain View, CA, USA\\
  $^2$Sensory, Portland, OR, USA\\
  $^3$Xiaomi Inc., Beijing, China}
\begin{document}
%\ninept
%
\maketitle
\begin{abstract}
This paper proposes \textit{multistream CNN}, a novel neural network architecture for robust acoustic modeling in speech recognition tasks. The proposed architecture processes input speech with diverse temporal resolutions by applying different dilation rates to convolutional neural networks across multiple streams to achieve the robustness. The dilation rates are selected from the multiples of a sub-sampling rate of 3 frames. Each stream stacks TDNN-F layers (a variant of 1D CNN), and output embedding vectors from the streams are concatenated then projected to the final layer. We validate the effectiveness of the proposed multistream CNN architecture by showing consistent improvements against Kaldi's best TDNN-F model across various data sets. Multistream CNN improves the WER of the test-other set in the LibriSpeech corpus by 12\% (relative). On custom data from ASAPP's production ASR system for a contact center, it records a relative WER improvement of 11\% for customer channel audio to prove its robustness to data in the wild. In terms of real-time factor, multistream CNN outperforms the baseline TDNN-F by 15\%, which also suggests its practicality on production systems. When combined with self-attentive SRU LM rescoring, multistream CNN contributes for ASAPP to achieve the best WER of 1.75\% on test-clean in LibriSpeech. 
\end{abstract}

\section{Introduction}

Automatic speech recognition (ASR) with processing speech inputs in multiple streams, namely \textit{multistream ASR}, has long been researched mostly for robust speech recognition tasks in noisy environments since the earlier works such as \cite{bourlard96-1,bourlard96-2,hynek96}. The multistream ASR framework was proposed based on the analysis of human perception and decoding of speech, where acoustic signals enter into the cochlea and are broken into multiple frequency bands such that the information in each band can be processed in parallel in the human brain \cite{allen94}. This approach worked reasonably well in the form of multi-band ASR where band-limited noises dominate signal corruption \cite{bourlard97,tibrewala97-2}. Later, further development was made in regards with multistream ASR in the areas of spectrum modulation and multi-resolution based feature processing \cite{hynek99,hynek05,tuske18} and stream fusion or combination \cite{okawa98,morris01,mesgarani11,mallidi15}.

With the advent of deep learning, multistream ASR research shifts its focus on deep neural network (DNN) architectures where multiple streams of encoders process embedding vectors in parallel. Although some forms of artificial neural networks like multilayer perceptron (MLP) \cite{bourlard93} had already been utilized in the literature for multistream ASR \cite{hynek96,mallidi15}, they were shallow and their usage was limited to fusing posterior outputs from a classifier in each stream. The recent DNN architectures for multistream ASR instead perform more unified functions, not only processing information in parallel but combining the information streams to classify all at once. In \cite{mallidi16-1,mallidi16-2} a multistream ASR architecture was simplified into one neural network where a binary switch was randomly applied to each feature stream when concatenating the multistream features as the neural network input. In decoding, a tree search algorithm was utilized to find the best stream combination. In \cite{li18}, a stream attention mechanism inspired by the hierarchical attention network \cite{yang16} was proposed to multi-encoder neural networks that can accommodate diverse viewpoints when processing embedding vectors. This multi-encoder architecture was successful in data sets recorded with multiple microphones \cite{barker17,ravanelli16}. As multi-head self-attention \cite{vaswani} became widely employed, multistream self-attention architectures were also investigated in \cite{han19-1,han19-2}. 

This paper presents \textit{multistream CNN} (as illustrated in Figure 1) as a novel neural network architecture for robust speech recognition. The proposed architecture processes input speech with diverse temporal resolutions by having stream-specific dilation rates to convolutional neural networks (CNNs) across multiple streams to achieve the robustness. In each stream we stack TDNN-F\footnote{TDNN-F stands for factorized time-delay neural network \cite{povey18tdnnf}. The convolution matrix in TDNN-F is decomposed into two factors with the orthonormal constraint, followed by a skip connection, batch normalization and a dropout layer.}, a variant of 1D-CNN. The dilation rate for the TDNN-F layers in each stream is chosen from multiples of the default sub-sampling rate (3 frames) for model training and decoding. The choice of multiples of 3 for the dilation rates can offer a seamless integration with the training and decoding process. Output embedding vectors from the streams are concatenated then projected to the final layer.

\begin{figure}[t]
  \centering
  \includegraphics[width=\linewidth]{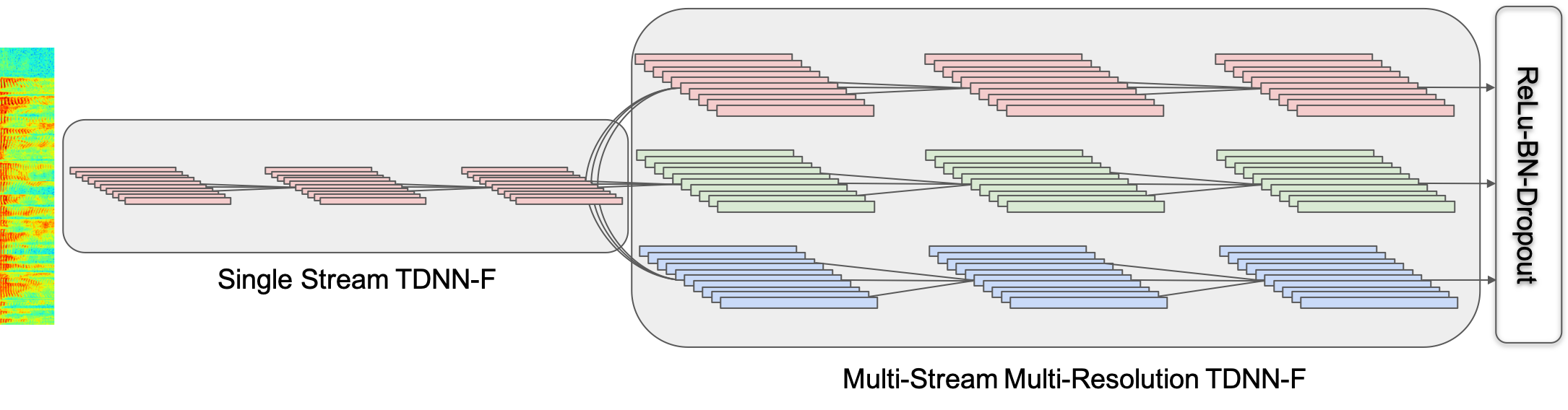}
  \caption{Schematic diagram of the proposed multistream CNN architecture.}
  \label{fig:speech_production}
  \vspace{-0.5cm}
\end{figure}

We structure this paper as follows. In Section 2, we detail training and evaluation data, and share experimental setups for ablation discussions in Section 3, where we explain our proposal of multistream CNN and analyze the impact of a few design choices in the proposed architecture on the LibriSpeech data. In Section 4, we discuss the performances of single- and multistream CNNs on custom data from ASAPP's production ASR system for a contact center in terms of both WER and RTF. In section 5, we conclude this work with summaries and comments on future directions. 

\section{Data and Experimental Setups}

\subsection{Data}

The LibriSpeech corpus \cite{panayotov2015librispeech} is a collection of approximately 1,000hr read speech (16kHz) from audio books. Each dev/test category (clean and other) contains around 5hrs of audio. This corpus also provides $n$-gram LMs trained on 800M token texts.

The Switchboard-1 Release 2 (LDC97S62) and Fisher English Training Part 1 and 2 (LDC2004S13, LDC2004T19, LDC2005S13, LDC2005T19) corpora total 2,000hrs of 8kHz telephony speech. We use the HUB5 eval2000 data (LDC2002S09, LDC2002T43) for evaluation.

We collect roughly 500hrs of 8kHz audio from our production ASR system. An eval set is 10hr of audio collection with a balanced distribution between agent and customer channel recordings.  

\subsection{Experimental Setups}

For LibriSpeech, neural networks for acoustic modeling are trained on the 960hr training set with the LF-MMI objective \cite{Povey16}. The learning rates are decayed from $10^{-3}$ to $10^{-5}$ over the span of 6 epochs. The minibatch size is 64. We use the $n$-gram LMs provided by the LibriSpeech corpus for the 1st pass decoding and 2nd pass rescoring. 

Regarding model training with SWBD/Fisher, we leverage the default Kaldi recipe\footnote{https://github.com/kaldi-asr/kaldi/tree/master/egs/fisher\_swbd/s5} \cite{Povey11}. We train models on the 2,000hr training data. For neural network AMs, we exponentially decay the learning rates from $10^{-3}$ to $10^{-4}$ during 6 epochs. The minibatch size is 128. The default $n$-gram LMs produced by the recipe are used for decoding.

We fine-tune the SWBD/Fisher model with the ASAPP custom data for further evaluation on data in the wild. We adjust the learning rate decay schedule, starting from $10^{-5}$ to $10^{-7}$ for 6 epochs with the minibatch size of 128. The PocoLM toolkit\footnote{https://github.com/danpovey/pocolm} is used to train a 4-gram LM for the 1st-pass decoding in the evaluation.   

\section{Multistream CNN}

As illustrated in Figure 1, the proposed multistream CNN architecture branches multiple streams after processing given input speech frames with 5 CNN layers in a single stream where CNNs could be TDNN-F or 2D-CNN (in case of applying SpecAugment \cite{specaugment}). After being branched out, a stack of 17 TDNN-F layers in each stream process the output of the single-streamed CNNs with a unique dilation rate. Consider an embedding vector $\mathbf{x}_i$ comes out of the single-streamed CNN layers at a given time step of $i$. An embedding vector $\mathbf{y}^m_i$ from a stream $m$ having gone through the stack of TDNN-F layers with a dilation rate $r_m$ can be written as below:
\begin{equation}
    \mathbf{y}^m_i = \textit{Stacked-TDNN-F}_m \left( \mathbf{x}_i ; \left[ -r_m, r_m \right] \right),
\end{equation}
where $[ -r_m, r_m]$ means a $3 \times 1$ kernel with the dilation rate $r_m$ for each TDNN-F layer. It is crucial to choose $r_m$ from the multiples of a sub-sampling rate used for model training and decoding. (In our case, we choose the multiples of 3 frames.) Output embedding vectors from all the streams are concatenated and followed by ReLu, batch normalization and a dropout layer;
\begin{equation}
    \mathbf{z}_i = \textit{Dropout} \left( \textit{BN} \left( \textit{ReLu} \left( \textit{Concat} \left( \mathbf{y}^1_i, \mathbf{y}^2_i, \dots, \mathbf{y}^M_i \right) \right) \right) \right),
\end{equation}
which is projected to the output layer via a couple of fully connected layers.

In next subsections, we analyze the effect of our design choices in the proposed multistream CNN architecture using the LibriSpeech dev and test sets. Unless specified, the complexity of all the models compared in the analysis is around 20M parameters for a fair comparison. The baseline model is a 17-layer (single-stream) TDNN-F in the recipe\footnote{https://github.com/kaldi-asr/kaldi/egs/librispeech/s5/local/chain/run\_tdnn.sh} for LibriSpeech of the Kaldi toolkit.

\subsection{Multiple Streams w/ Dilation Rates}

Table 1 compares multistream CNN models against the baseline as we increase the number of streams with various dilation rates. For example, \texttt{1-2-3-4-5} indicates that the dilation rates of 1, 2, 3, 4, and 5 are used to TDNN-F layers over the total 5 streams respectively in a multistream CNN model. We adjust the dimension of embedding vectors for the TDNN-Fs to keep the model complexity around 20M parameters for a fair comparison. From the upper half of the table, it is evident that the proposed multistream CNN architecture improves the WERs of the `other' data sets more noticeably as we increase the number of streams up to 9 by the increment of 1. We don't report model performances with more streams since we observed no improvement after 9 streams. This could have resulted from combined reasons, such as too small embedding dimension for TDNN-F over too many streams. In the lower half of the table, we apply the multiples of the sub-sampling rate (i.e., 3 frames). The results prove careful selection of dilation rates would further improve WER even with smaller numbers of streams. The choice of the multiples of 3 for TDNN-F layers seems to be better streamlined with the training and decoding process where input speech frames are sub-sampled every 3 frames. 

\begin{table}[t]
\centering
    \caption{LibriSpeech WERs (\%) by multistream CNNs with various combinations of dilation rates across streams. $d$: embedding dimension for TDNN-F.}
    \renewcommand{\arraystretch}{1.25}
    \begin{tabular}{C{2cm}|C{0.7cm}|C{0.7cm}|C{0.7cm}|C{0.7cm}|C{0.7cm}}
        \hline
        \centering \multirow{2}{*}{\small System} & \multirow{2}{*}{\small $d$} & \multicolumn{2}{c|}{\small dev} & \multicolumn{2}{c}{\small test}\\
        \cline{3-6} 
        \centering & & \small clean &  \small  other & \small clean & \small other \\
        \hline
        \hline
        \centering \small Baseline & \small 1,536 & \small 3.26 & \small 8.86 & \small 3.68 & \small 8.92 \\
        \hline
        \hline
        \multicolumn{6}{c}{Multistream CNN} \\
        \hline
        \centering \small 1-2 & \small 768 & \small 3.36 & \small 8.86 & \small 3.80  & \small 8.91 \\
        \hline
        \centering \small 1-2-3 & \small 512 & \small 3.33 & \small 8.81 & \small 3.84  & \small 8.93 \\
        \hline
        \centering \small 1-2-3-4-5 & \small 307 & \small 3.36 & \small 8.62 & \small 3.67 & \small 8.76 \\
        \hline
        \centering \small 1-2- $\cdots$ -6-7 & \small 219 & \small 3.27 & \small \textbf{8.39} & \small 3.65 & \small 8.85 \\
        \hline
        \centering \small 1-2- $\cdots$ -8-9 & \small 170 & \small 3.27 & \small 8.45 & \small \textbf{3.60} & \small \textbf{8.63} \\
        \hline
        \hline
        \centering \small 1-3-6 & \small 512 & \small 3.27 & \small 8.58 & \small 3.67  & \small 8.71 \\
        \hline
        \centering \small 3-6-9 & \small 512 & \small 3.24 & \small 8.30 & \small 3.59  & \small 8.70 \\
        \hline
        \centering \small 6-9-12 & \small 512 & \small \textbf{3.17} & \small \textbf{8.25} & \small \textbf{3.54} & \small \textbf{8.41} \\
        \hline
        \centering \small 1-3-6-9-12 & \small 307 & \small 3.29 & \small 8.26 & \small 3.57 & \small 8.78 \\
        \hline
        \centering \small 3-6-9-12-15 & \small 307 & \small 3.22 & \small 8.30 & \small 3.58 & \small 8.52 \\
        \hline
    \end{tabular}
    %\vspace{-0.5cm}
    \label{tab:adaptation}
\end{table}

We find the best setup from the \texttt{6-9-12} configuration, which, juxtaposed with the baseline, shows a relative WER improvements of 6.9\% and 5.7\% on dev-other and test-other, respectively. 

%The diversity of streams in terms of temporal resolution seems another critical factor for a multistream CNN architecture to achieve the expected performance. Comparing the WERs of the models with the \texttt{1-3-6-9-12-15} and \texttt{(1-3-6)}$^2$ configuration\footnote{\texttt{(1-3-6)}$^2$ equals \texttt{1-3-6-1-3-6}, and \texttt{(1-3-6)} repeats 2 times over 6 streams.}, the model with unique dilation rates in each stream is shown to be superior to the model otherwise. A similar observation can be made by considering the WERs of the multistream CNN model with the \texttt{1-2-3-4-5-6-7-8-9} configuration in Table 1 and those with the configurations of \texttt{(1-3-6)}$^3$, \texttt{(3-6-9)}$^3$ and \texttt{(6-9-12)}$^3$ in Table 2, all of which have 9 streams in the model architecture. Without stream diversity in terms of temporal resolution, it is recognized that multistream CNNs configured with dilation rates from multiples of 3 would not outperform those not configured with multiples of 3. 

%We find the best setup from the \texttt{6-9-12} configuration, except for the dev-other set where the \texttt{(6-9-12)}$^3$ outperforms other configurations.

\begin{table}[t]
\centering
    \caption{LibriSpeech WERs (\%) by multistream CNNs in larger size. $N$: model complexity in \# of parameters. $d$: embedding dimension for TDNN-F.}
    \renewcommand{\arraystretch}{1.25}
    \begin{tabular}{C{1.2cm}|C{0.8cm}|C{0.7cm}|C{0.6cm}|C{0.6cm}|C{0.6cm}|C{0.6cm}}
        \hline
        \centering \multirow{2}{*}{\small System} & \multirow{2}{*}{\small $N$} & \multirow{2}{*}{\small $d$} & \multicolumn{2}{c|}{\small dev} & \multicolumn{2}{c}{\small test}\\
        \cline{4-7} 
        \centering & & & \small clean &  \small  other & \small clean & \small other \\
        \hline
        \hline
        \centering \small Baseline& \small 20.7M & \small 1,536 & \small 3.26 & \small 8.86 & \small 3.68 & \small 8.92 \\
        \hline
        \hline
        \multicolumn{7}{c}{Multistream CNN} \\
        \hline
        \centering \small 6-9-12 & 20.6M & \small 512 & \small 3.17 & \small 8.25 & \small 3.54  & \small 8.41 \\
        \hline
        \centering \small 6-9-12 & 73.2M & \small 1,536 & \small \textbf{3.07} & \small 8.10 & \small \textbf{3.40}  & \small \textbf{8.32} \\
        \hline
        \centering \small 6-9-12 & 93.9M & \small 1,536 & \small 3.09 & \small \textbf{7.98} & \small 3.52 & \small \textbf{8.32} \\
        \hline
    \end{tabular}
    %\vspace{-0.5cm}
    \label{tab:adaptation}
\end{table}

\subsection{Larger Networks}
Table 2 contrasts the WERs of the multistream CNN models with the same \texttt{6-9-12} configuration, but with different model complexity. The 73M parameter model has 3 times larger embedding dimension for TDNN-F (1,536 versus 512), while the 94M parameter model has 7 more TDNN-F layers in each stream. As observed in the table, the larger-sized multistream CNN models reached lower WERs, but the improvement from the 20M parameter model seems marginal considering much longer training times. In real-world applications, especially for cases where online inference is critical, the 20M parameter model must be a reasonable choice.   

\subsection{Toward State-of-the-Art}
In this section, we optimize the multistream CNN model with the \texttt{6-9-12} configuration (20M parameter) with SpecAugment and neural network based LMs toward competitive state-of-the-art results in LibriSpeech.

Since its introduction in \cite{specaugment}, the SpecAugment data augmentation method of masking random time-frequency bands from input spectrograms has been wildly adopted by both hybrid and end-to-end ASR systems. SpecAugment is known to prevent neural network models from being overfit thus enable them to become more robust to unseen testing data. To apply this method on top of the proposed multistream CNN architecture, we replace the first 5 layers of TDNN-F of the model (corresponding to the Single Stream TDNN-F part in Figure 1) with 5 layers of 2D-CNN to better accommodate log-mel spectrograms. We use $3 \times 3$ kernels for the 2D-CNN layers with a filter size of 256 except for the first layer with the filter size of 128. Every other layer we apply frequency band sub-sampling with the rate of 2. 

We employ multiple stages of LM rescoring in order to obtain the minimum WERs on the test sets in LibriSpeech. The LMs are trained on normalized texts where typos are corrected as well as spelling consistencies between British and American English are addressed. The initial decoding is based on the decoding graph constructed from the multistream CNN AM and a 3-gram LM, resulting in the initial hypotheses in a lattice format. Lattice rescoring is done with a larger sized 4-gram LM, followed by a second-pass lattice rescoring with a TDNN-LSTM language model \cite{Li2018}. In the final rescoring stage, we use an interpolated self-attentive SRU LM \cite{pan2020}. We linearly interpolate two self-attentive SRU models, one of which is trained on word pieces using byte-pair encoding (BPE) and the other is trained at a word level. After the interpolation, we re-rank the $N$-best hypotheses from the lattices rescored by the TDNN-LSTM LM in the previous stage. In our experiments, we empirically keep $N$ at 100. 

In Table 3, we tabulate the performances of the three LM rescoring stages for multistream CNN where the 4-gram LM rescores the first pass decoding results (first line) and the other two neural network based LMs further rescore the n-gram LM rescored results. The 1.75\% WER on test-clean is, to the best of the authors' knowledge, the lowest reported in the literature, without extra data (e.g., Libri-Light) taken into consideration for either AM or LM training.  

\begin{table}[t]
\centering
\caption{State-of-the-art performances on LibriSpeech using multistream CNN and neural network based LMs including self-attentive SRU LM.}
\renewcommand{\arraystretch}{1.25}
\begin{tabular}{L{2.7cm}|C{0.7cm}|C{0.7cm}|C{0.7cm}|C{0.7cm}}
\hline
\hspace{8mm} \multirow{2}{*}{\small Setup} & \multicolumn{2}{c|}{\small dev} & \multicolumn{2}{c}{\small test} \\
\cline{2-5} 
& \small clean & \small other & \small clean & \small other \\
\hline
\small Multistream CNN & \small 2.62 & \small 6.78 & \small 2.80 & \small 7.06 \\ 
\hline
\small +TDNN-LSTM LM & \small 2.14 & \small 5.82 & \small 2.34 & \small 6.04 \\
\hline
\small +Self-Attentive SRU & \small 1.55 & \small 4.22 & \small \textbf{1.75} & \small 4.46 \\
\hline
\end{tabular}
%\vspace{-0.5cm}
\label{table:wer_setup}
\end{table}

\section{Multistream CNN in the Wild}
In this section, we manifest the feasibility of the proposed multistream CNN architecture in real-world scenarios. We use our custom training data (500hrs) collected from ASAPP's contact center ASR system to fine-tune the seed models (baseline TDNN-F and multistream CNN) trained on the SWBD/Fisher corpora mentioned in Section 2.1. The seed model performances on the HUB5 eval2000 data consisting of the SWBD and CH (i.e., CallHome) portions are presented in Table 4. A noteworthy observation in the table is that the proposed model architecture continues to excel the baseline model in more challenging data. This is further highlighted in Table 5 where the two fine-tuned models (baseline and multistream CNN) are evaluated on the ASAPP custom eval set of 10hrs also described in Section 2.1. The relative WER improvement (WERR) of 11.4\% on the customer channel recordings\footnote{Compared to agent channel audio, customer channel audio are by far challenging for ASR systems due to noisier acoustic environments, non-native/accented speech, multiple talkers, etc.} declares the robustness of the proposed multistream CNN model architecture in the wild. In addition, the relative real-time factor (RTF) improvement of 15.1\% against the baseline TDNN-F model shows the practicality of the proposed model architecture in real-world applications, especially where online inference is necessary. 

\begin{table}[t]
\centering
    \caption{HUB5 eval2000 WERs (\%) by telephony seed models. SWBD: Switchboard, CH: CallHome in HUB5 eval2000.}
    \renewcommand{\arraystretch}{1.25}
    \begin{tabular}{C{1.5cm}|C{1.1cm}|C{1.1cm}|C{1.1cm}|C{1.1cm}}
        \hline
        \centering \multirow{2}{*}{} & \multicolumn{2}{c|}{\small Baseline} & \multicolumn{2}{c}{\small Multistream CNN}\\
        \cline{2-5} 
        \centering & \small SWBD &  \small  CH & \small SWBD & \small CH \\
        \hline
        \hline
        \centering \small WER  (\%)& \small 8.7 & \small 16.2 & \small 9.0 & \small 15.6 \\
        \hline
    \end{tabular}
    %\vspace{-0.25cm}
    \label{tab:adaptation}
\end{table}

\begin{table}[t]
\centering
    \caption{Relative performance improvements by multistream CNN on ASAPP's custom data for conversational speech over telephony channels, against the baseline TDNN-F model. The absolute performances are not disclosed. RTF: real time factor.}
    \renewcommand{\arraystretch}{1.25}
    \begin{tabular}{C{2.5cm}|C{1.25cm}|C{1.25cm}|C{1.5cm}}
        \hline
        \centering \multirow{2}{*}{} & \multicolumn{2}{c|}{\small WERR (\%)} & Relative \\
        \cline{2-3}
        \centering  & \small  Agent & \small Customer & RTF Imp. \\
        \hline
        \hline
        \centering \small Multistrem CNN & \small 8.8 & \small 11.4 & \small 15.1 \\
        \hline
    \end{tabular}
    %\vspace{-0.25cm}
    \label{tab:adaptation}
\end{table}

\section{Conclusions}
In this paper, we proposed a novel neural network architecture, namely multistream CNN, for robust speech recognition. The reasoning behind the proposal was that diversity in temporal resolution across multiple streams would enhance the overall robustness in acoustic modeling. We empirically showed that it would further improve the benefit of having such diversity in temporal resolution to choose dilation rates for TDNN-F layers across multiple streams form the multiples of 3 frames (i.e., sub-sampling rate). We tested multistream CNN models on various data sets including ASAPP's custom data collected from a contact center ASR system to demonstrate the robustness and practicality of the proposed model architecture.  

Multistream CNN seems promising to be utilized in a number of ASR applications and frameworks. We plan to continue to improve this multistream model architecture to further enhance our production ASR systems.

%\vfill\pagebreak

% References should be produced using the bibtex program from suitable
% BiBTeX files (here: strings, refs, manuals). The IEEEbib.bst bibliography
% style file from IEEE produces unsorted bibliography list.
% -------------------------------------------------------------------------
\bibliographystyle{IEEEbib}

\begingroup
\setstretch{0.9}
%\printbibliography

\bibliography{strings,refs}
\endgroup
\end{document}